\documentclass[pra,aps,twocolumn,showpacs]{revtex4}
\usepackage{epsfig,amsfonts,amsmath,amssymb}
\newcommand{\ket}[1]{|#1\rangle}
\newcommand{\bra}[1]{\langle #1|}
\newcommand{\proj}[1]{\ket{#1}\bra{#1}}
\begin{document}

\title{Entanglement in a class of multiqubit mixed states
without multipartite tangles}

\author{Yan-Kui Bai,$^{1,2}$ Ming-Yong Ye,$^{1,3}$ and Z. D. Wang$^1$}
\email{zwang@hkucc.hku.hk}
 \affiliation{$^1$ Department of Physics and
Center of Theoretical and Computational Physics, University of Hong
Kong, Pokfulam Road, Hong Kong, People's Republic of China\\$^2$
College of Physical Science and Information Engineering $\&$ Hebei
Advance Thin Films Laboratory, Hebei Normal University,
Shijiazhuang, Hebei 050016, People's Republic of China\\$^3$School
of physics and Optoelectronics Technology, Fujian Normal University,
Fuzhou 350007, People's Republic of China}

\begin{abstract}

Based on quantum complementary relations (QCRs) and a purification
scenario, we analyze a class of N-qubit mixed states that are
entangled but do not have two-, and genuine three-, four-, ...,
N-qubit entanglements. It is shown that entanglement (one-tangle or
negativity) in these mixed states is closely related to the QCR
entanglement of their purified states. In particular, it is
elaborated that when the mixed state does not have multipartite
tangles (two- and higher tangles), its entanglement is actually a
kind of genuine multipartite QCR entanglement between the system and
its environment.

\end{abstract}


\maketitle

Entanglement plays a crucial role in quantum-information processing,
including quantum communication \cite{eke91,ben92,ben93} and quantum
computation \cite{ben00,rau01,llb01}. Therefore, it is highly
desirable and necessary to characterize the entanglement property of
quantum systems. So far, only bipartite entanglement has been
understood well in many aspects \cite{rev07}, while the
characterization of multipartite entanglement for quantum many-body
systems, especially for their mixed states, has still been very
challenging despite a number of profound results, e.g.,
\cite{lrefs,arefs}.

Uhlmann worked out that mixed-state entanglement measure may be
constructed by the convex roof extension of pure-state measure
\cite{uhl98}. For example, when an entanglement measure $E(\psi)$ is
available for pure states, the corresponding measure for a given
mixed state can be expressed as
\begin{equation}\label{1}
    E(\rho)=\mbox{min}\sum p_i E(\psi_{i}),
\end{equation}
where the minimum runs over all the pure-state decomposition $\{p_i,
\psi_i \}$ of $\rho$. As a result, an analytical entanglement
formula of any two-qubit system, i.e., the concurrence $C(\rho)$,
was derived \cite{woo97}, and has been widely accepted, though the
analytical results for generic multiqubit systems can hardly be
obtained.

As is known, entanglement of a multi-qubit pure state has likely a
hierarchy structure: it is contributed from different levels of
quantum correlation. A quantitative relation for a three-qubit pure
state is given by Coffman, Kundu, and Wootters (CKW) \cite{ckw00}
$\tau_{A}(\psi_{ABC})=C^2(\rho_{AB})+C^2(\rho_{AC})+\tau_3(\psi_{ABC}),$
where the linear entropy (one-tangle) $\tau_A$ \cite{san00}
quantifies the bipartite entanglement in partition $A|BC$, and the
square of concurrence $C^2$ (two-tangle) and three-tangle $\tau_3$
\cite{dur00} quantify two- and genuine three-qubit entanglement,
respectively. Recently, Lohmayer \emph{et al} \cite{loh06} analyzed
the CKW relation for a kind of mixed state $\rho_{ABC}$, in which
the bipartite entanglement $\tau_{A}$ is nonzero while no two- and
three-qubit entanglement is present. This is remarkably different
from that in the pure state cases. More interestingly, we even find
below that this feature exists for a class of $N$-qubit ($N>2$)
mixed states, \emph{i.e.}, the mixed state is entangled but all the
two-, three-, ... , and N- tangles vanish. Thus it is natural to ask
\emph{where this kind of entanglement comes from}.

A mixed state can be interpreted as the partial trace of a larger
pure state composed of the concerned system and its environment
\cite{coh97}. In this sense, by analyzing the pure state, one may
obtain more valuable information about the mixed state.

In this paper, motivated by the above question, we analyze in detail
the entanglement in this class of $N$-qubit mixed states without
multipartite tangles. It is shown that this kind of mixed state
entanglement is closely related to their purified states. In
particular, it is elaborated that the one-tangle (or negativity
\cite{gvi02}) in this class of mixed states is a kind  of genuine
multipartite QCR-entanglement between the system and its
environment, focusing on (i) the three-qubit mixed state addressed
by Lohmayer \emph{et al} before; (ii) a new four-qubit case, namely
the one-parameter Smolin state \cite{smo01}; and (iii) a more
general $N$-qubit case.

\emph{The QCRs and a multipartite entanglement measure--} As a
fundamental principle, the QCR is often referred to the mutually
exclusive properties of a single quantum system. A quantitative
version in an N-qubit pure state $\ket{\Psi_{N}}$ is \cite{qcr05}
\begin{equation}\label{2}
    \tau_{k(R_k)}+S^{2}_{k}=1,
\end{equation}
where the linear entropy $\tau_{k(R_k)}=2(1-\mbox{tr}\rho_k^2)$
\cite{san00} characterizes the bipartite quantum entanglement
between qubit $k$ and the remaining qubits $R_k$, and $S^{2}_{k}$ is
a measure of single-particle property. Based on this relation, a
multipartite QCR-entanglement measure is introduced \cite{byw07},
\begin{equation}\label{3}
    E_{ms}(\Psi_{N})=\frac{\sum_{k} \tau_{k(R_k)}-2\sum_{i<j}
C_{ij}^2}{N},
\end{equation}
which is used to characterize total multi-qubit entanglement,
\emph{i.e.}, the sum of genuine three-, four-, ... , N-qubit
entanglements in the pure state. Here, the two-qubit entanglement in
$\ket{\Psi_N}$ is quantified by the square of the concurrence which
is defined as $C_{ij}=\mbox{max}[0,
(\sqrt{\lambda_1}-\sqrt{\lambda_2}-\sqrt{\lambda_3}-\sqrt{\lambda_4})]$
with the decreasing positive real numbers $\lambda_{i}$ being the
eigenvalues of the matrix
$\rho_{ij}(\sigma_y\otimes\sigma_y)\rho_{ij}^{\ast}(\sigma_y\otimes\sigma_y)$
\cite{woo97}. In a three-qubit pure state, $E_{ms}$ characterizes
genuine tripartite entanglement and is just the three-tangle
$\tau_3$. For the four-qubit case, $E_{ms}$ quantifies the sum of
genuine three- and four-qubit entanglements. The result of much
numerical analysis agrees with the conjecture that $E_{ms}$ is
entanglement monotone \cite{byw07}. Especially, for four-qubit
cluster-class states, an analytical proof is given and a set of
hierarchy entanglement measures is obtained \cite{bai08,ren08}. In a
general $N$-qubit pure state, $E_{ms}$ is conjectured to be
entanglement monotone. In Eq.(3), when the concurrences are zero, it
is obvious that the property holds.

\begin{figure}
\begin{center}
\epsfig{figure=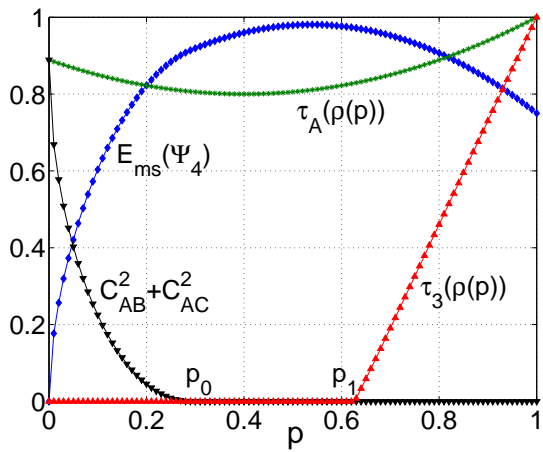,width=0.4\textwidth}
\end{center}
\caption{(Color online) The average multipartite entanglement
$E_{ms}$ (blue line) versus the decomposition parameter $p$ in the
purified state $\ket{\Psi_4}$, in comparison with the concurrence
(black line), the three-tangle (red line), and the one-tangle (green
line) of the state $\rho(p)$ plotted in Ref. \cite{loh06}.}
\end{figure}

\emph{Entangled three-qubit mixed states--} Lohmayer, \emph{et al.}
considered the following mixed state \cite{loh06}
\begin{equation}\label{4}
    \rho_{ABC}(p)=p \proj{GHZ}+(1-p) \proj{W},
\end{equation}
where the real parameter $p$ ranges in $[0,1]$, and the two
orthogonal pure states have forms
$\ket{GHZ}=(\ket{000}+\ket{111})/\sqrt{2}$ and
$\ket{W}=(\ket{001}+\ket{010}+\ket{100})/\sqrt{3}$, respectively. As
shown in Fig.1, the quantum state is entangled ($\tau_{A}\neq 0$) in
the region $[p_0,p_1]$, but the entanglement is neither two-qubit
entanglement nor genuine three-qubit entanglement (here $p_0\approx
0.2918$ and $p_1\approx 0.6269$).

With the reduction interpretation of mixed states, we consider a
pure state $\ket{\Psi}$, which satisfies
$\mbox{tr}_{\mathcal{E}}(\proj{\Psi})=\rho_{ABC}(p)$ with
$\mathcal{E}$ being environment system. Because $\rho_{ABC}(p)$ is
a rank-2 quantum state, it is sufficient to consider an
environment of dimension-2 according to the purification theorem
(a corollary of Schmidt decomposition) \cite{per93}. Therefore,
the environment system $\mathcal{E}$ is equivalent to a qubit. For
simplification, we consider the pure state
\begin{equation}\label{5}
    \ket{\Psi_4}=\sqrt{1-p}\ket{W}_{ABC}
    \ket{0}_{D}+\sqrt{p}\ket{GHZ}_{ABC}\ket{1}_{D},
\end{equation}
where the environment is represented by qubit $D$. Any other
purified state is equivalent to the $\ket{\Psi_4}$ plus a local
unitary transformation $U_D$, which does not change the
entanglement.

\begin{figure}[b]
\begin{center}
\epsfig{figure=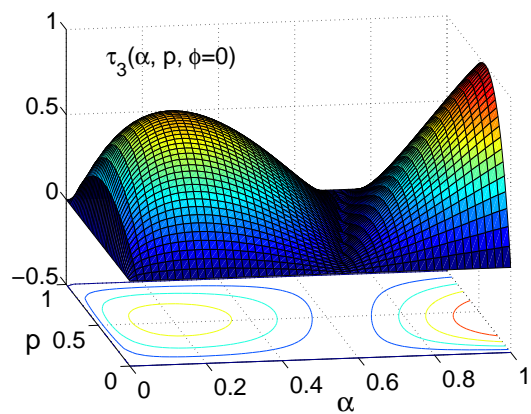,width=0.4\textwidth}
\end{center}
\caption{(Color online) The pure state three-tangle
$\tau_3(\Phi_{ABD})$ versus the parameters $p$ and $\alpha$ for the
relative phase $\phi=0$.}
\end{figure}

Since $\rho_{ABC}$ is a mixed state, the purified state
$\ket{\Psi_4}$ is bipartite entangled in the partition $ABC|D$. We
first analyze two-qubit entanglement in the pure state. Due to the
permutation invariance of qubits A, B, and C, there are only two
independent two-qubit reduced density matrices, \emph{i.e.},
$\rho_{AB}$ and $\rho_{AD}$. Entanglement in subsystem $AB$ is
$C_{AB}^2=(\mbox{max}[0,\frac{2}{3}(1-p)-\sqrt{\frac{p}{3}(2+p)}])^2$
\cite{loh06}. For the quantum state $\rho_{AD}$, we have
$C_{AD}^2=0$. In three-qubit reduced density matrices, only
$\rho_{ABC}$ and $\rho_{ABD}$ are independent. An analytical formula
of three-tangle for the quantum state $\rho_{ABC}(p)$ was given in
Ref. \cite{loh06}. However, the case for mixed state $\rho_{ABD}$ is
different; it has the form
\begin{equation}\label{6}
    \rho_{ABD}(\alpha_p,p)=\alpha_p\psi^{(1)}(p)+(1-\alpha_p)\psi^{(2)}(p),
\end{equation}
where $\psi^{(i)}=\proj{\psi^{(i)}}$ is a projector with
$\ket{\psi^{(1)}(p)}=\sqrt{1-a}\ket{000}+ \sqrt{a}\ket{111}$ and
$\ket{\psi^{(2)}(p)}=\sqrt{b}\ket{001}+
\sqrt{(1-b)/2}(\ket{010}+\ket{100})$, and the coefficients are
$\alpha_p=(2+p)/6$, $a=3p/(2-p)$ and $b=3p/(4-p)$, respectively.
Eltschka, \emph{et al} obtained an expression of three-tangle for
this kind of mixed states \cite{elt07}. We will show below that this
mixed state three-tangle is equal to zero. In general, the
pure-state component of $\rho_{ABD}$ can be written as
$\ket{\Phi}_{ABD}=\sqrt{\alpha}\ket{\psi^{(1)}(p)}
-e^{i\phi}\sqrt{1-\alpha}\ket{\psi^{(2)}(p)}$, where real number
$\alpha$ ranges in $[0,1]$ and relative phase $\phi$ in $[0,2\pi]$.
Its pure state three-tangle is
\begin{equation}\label{7}
   \tau_3(\alpha, p, \phi)=4|f_1(\alpha,p)-e^{i3\phi}f_2(\alpha,p)|,
\end{equation}
where $f_1=6\alpha p(1-p)/(2+p)^2$ and $f_2=24(p-p^2)/(4-p)
[\alpha(1-\alpha)^3/(8+2p-p^2)]^{1/2}$, respectively. When the
relative phase $\phi=2k\pi/3$, given a value of parameter $p$, there
always exists a nontrivial zero point
$\alpha_0(p)=[1+(2\sqrt[3]{2})^{-1}(6/(2+p)-1)]^{-1}$ at which the
pure state three-tangle is zero (in Fig.2, the $\tau_3(\Phi_{ABD})$
is plotted as functions of the parameters $\alpha$ and $p$, where
the relative phase $\phi=0$ is chosen). The cases for relative
phases $\phi=2\pi/3$ and $\phi=4\pi/3$ are the same due to the phase
factor $e^{3i\phi}$ in Eq.(7). Therefore, the $\tau_3$ for mixed
state $\rho_{ABD}(\alpha_0(p),p)$ is zero and its optimal
decomposition is $\{\Phi_k(2k\pi/3)\}$ with probabilities $p_k=1/3$
for $k=1, 2, 3$. According to the convex characteristic manifold
\cite{ost08}, we have $\tau_3(\rho_{ABD}(\alpha,p))=0$ when
$\alpha<\alpha_0(p)$, which is because the quantum state can be
decomposed into the mix of $\rho_{ABD}(\alpha_0,p)$ and
$\psi_{ABD}^{(2)}$ (a $W$-class state). In Eq. (6), the parameter
$\alpha_p=(2+p)/6$ ranges in $[1/3,1/2]$ which is less than the
$\alpha_0(p)\in [0.5575, 0.7159]$, therefore $\tau_3(\rho_{ABD})=0$.
In addition, we have $\tau_3(\rho_{ACD})=\tau_3(\rho_{BCD})=0$ in
terms of the permutation invariance.

For the whole pure state $\ket{\Psi_4}$ in Eq. (5), we can derive
the multipartite QCR entanglement
\begin{eqnarray}\label{8}
  E_{ms}^{I}(p\leq p_0) &=& \frac{3p(2-3p)}{4}+\frac{2(1-p)
                    \sqrt{p(2-p)}}{\sqrt{3}}, \nonumber\\
  E_{ms}^{II}(p>p_0) &=& \frac{8+(14-13p)p}{12},
\end{eqnarray}
which characterizes the genuine three- and four-qubit entanglements.
As shown in Fig.1, we plot the variation of the measure along with
parameter $p$. In the region $[0,p_1]$, due to all the three-qubit
entanglement $\tau_3(\rho_{ijk})$s being zero, the $E_{ms}$
quantifies only the genuine four-qubit entanglement $\tau_4$ which
attains to the maximum $0.9808$ at $p=7/13$. When $p_1<p\leq1$,
though the tripartite entanglement $\tau_3(\rho_{ABC})$ is
increasing, the multipartite entanglement $E_{ms}$ decreases with
the parameter $p$. This is because the decrease of $\tau_4$ is
stronger than the increase of $\tau_3$.

The mixed state one-tangle $\tau_{A}(\rho_{ABC})$ is nonzero in the
whole region (green line in Fig.1), which means the subsystems $A$
and $BC$ are always entangled. However, in the region $[p_0, p_1]$,
the entanglement is neither two-qubit nor genuine three-qubit
entanglement \cite{loh06}. Our understanding lies in the fact that
here \emph{the entanglement comes from a kind of genuine four-qubit
QCR entanglement of the system $ABC$ and its environment $D$ since
all the $2$-, and $3$-tangles are zero in the purified state
$\ket{\Psi_4}$}. The genuine four-qubit entanglement can reduce by a
positive operator value measure (POVM) on qubit $D$. The one-tangle
$\tau_{A}(\rho_{ABC})=\mbox{min}\sum p_i\tau_A(\psi_{ABC}^{i})$
characterizes the minimal average entanglement in the partition
$A|BC$, for which its pure state component in the optimal
decomposition has the form
$\ket{\psi(\varphi_k)}=\sqrt{p}\ket{GHZ}-e^{i\varphi_k}\sqrt{1-p}\ket{W}$.
A two-component decomposition $\{\varphi_1=0, \varphi_2=\pi\}$
corresponds to a projection measure on environment $D$. In this
case, the genuine four-qubit entanglement $\tau_4(\Psi_4)$ reduces
to bipartite entanglement as measured by the one-tangle. The case
for the decomposition $\{\varphi_1=0, \varphi_2=\frac{2\pi}{3},
\varphi_3=\frac{4\pi}{3}\}$ is similar and related to a
three-outcome POVM.

\begin{figure}[t]
\begin{center}
\epsfig{figure=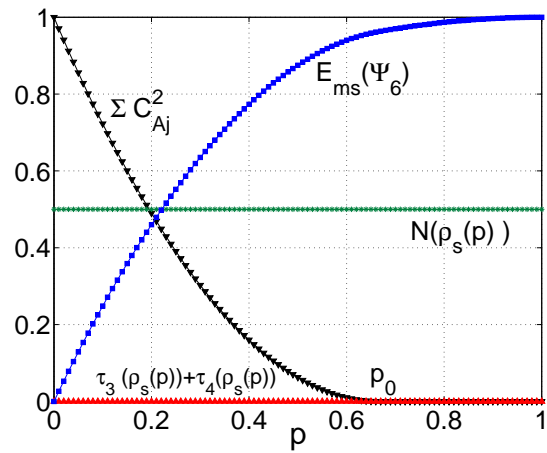,width=0.4\textwidth}
\end{center}
\caption{(Color online) The average multipartite entanglement
$E_{ms}(\Psi_6)$ (blue line), average negativity
$\mathcal{N}(\rho_S(p))$ (green line), concurrence sum $\sum
C_{Aj}^{2}$ (black line), and the sum of three- and four-tangle
$\tau_3(\rho_S(p)) +\tau_4(\rho_S(p))$ (red line) as a function of
$p$.}
\end{figure}

\emph{Entangled four-qubit mixed states--}The one-parameter Smolin
state is
\begin{equation}\label{9}
    \rho_S(p)=\frac{p}{4}\sum_{i=1}^{3} \psi^{(i)}_{AB}\otimes\psi^{(i)}_{CD}
    +(1-\frac{3p}{4})\psi_{AB}^{(4)}\otimes\psi_{CD}^{(4)},
\end{equation}
where $\psi^{(i)}=\proj{\psi^{(i)}}$ is the projector on Bell basis
$\{\ket{\Phi^{\pm}}, \ket{\Psi^{\pm}}\}$ \cite{smo01}. The quantum
state has the symmetry of qubit permutations $A\leftrightarrow B$,
$C\leftrightarrow D$, and $AB\leftrightarrow CD$. When the parameter
$p=1$, $\rho_S$ is the original Smolin state \cite{aug06, hil99}
which can maximally violate Bell inequalities and lead to secure
quantum secret sharing. In the mixed state $\rho_S(p)$, there are
only two independent two-qubit reduced density matrices, for which
the entanglements are $C_{AB}^2=[\mbox{max}(0, 1-3p/2)]^2$ and
$C_{AC}^2=0$, respectively. For the three-qubit reduced density
matrix $\rho_{ijk}$, its pure state component can be written as
$\ket{\mu}_{ij}\otimes\ket{\nu}_k$, so the three-tangle
$\tau_3(\rho_{ijk})=0$. The pure state component of $\rho_S(p)$ is
the tensor product state of two Bell states. According to the
formula $\tau_4(\rho)=\mbox{min}\sum p_i \tau_4(\psi_{i}^c)$
\cite{bai08}, we can derive the four-tangle $\tau_4(\rho_S(p))=0$.
In Fig.3, we plot the concurrence, three- and four-tangles as a
function of the parameter $p$. It is found that  all entanglements
are zero in the region $[p_0,1]$ (where $p_0=2/3$). However, the
negativity \cite{gvi02}
$\mathcal{N}_{A}(\rho_S(p))=(||\rho_S^{T_A}(p)||-1)/2$ is nonzero in
the region (the norm is defined as
$||\sigma||=\mbox{tr}\sqrt{\sigma\sigma^\dagger}$), which means that
bipartite entanglement exists in the partition $A|BCD$ (due to the
computational complexity, we do not choose the mixed-state
one-tangle). The cases for partitions $B|ACD$, $C|ABD$ and $D|ABC$
are similar, and we plot the average negativity
$\mathcal{N}(\rho_S(p))=\sum \mathcal{N}_{k}(\rho_S(p))/4$ in Fig.3.
In analogy to the three-qubit case, we show here that the negativity
is only connected to a kind of genuine multipartite QCR entanglement
between the system and its environment.

According to the purification theorem, it is sufficient to consider
an environment $\mathcal{E}$ of dimension-4. Therefore, the
$\mathcal{E}$ is equivalent to two qubits and the purified state of
$\rho_{S}(p)$ can be written as
\begin{equation}\label{10}
    \ket{\Psi_6(p)}=\sum_{i,j=0,1} \xi_{ij}\ket{\psi^{(ij)}}_{AB}
    \ket{\psi^{(ij)}}_{CD}\otimes \ket{ij}_{EF},
\end{equation}
where the $\ket{\psi^{(ij)}}$ is the Bell state, and the
coefficients are $\xi_{00}=\xi_{01}=\xi_{10}=\sqrt{p/4}$ and
$\xi_{11}=\sqrt{1-3p/4}$. For this pure state, the multipartite QCR
entanglement $E_{ms}(\Psi_6)$ characterizes the sum of genuine
three-, four-, five-, and six-qubit entanglements. After some
derivation, we have $E_{ms}(p\leq p_0)=5p(1-p)/3$ and
$E_{ms}(p>p_0)=(2+2p-p^2)/3$. In the quantum state $\ket{\Psi_6}$,
the pure-state component of reduced density matrices $\rho_{ijk}$
and $\rho_{ijkl}$ can be written as a tensor product state.
Therefore the mixed-state three- and four-tangles are zero in the
purified state, and the $E_{ms}$ quantifies only genuine five- and
six-qubit QCR-entanglement. After a POVM on the environment system
$EF$, the entanglement $E_{ms}(\Psi_6)$ reduces to the bipartite
entanglement in the one-parameter Smolin state as measured by the
negativity $N(\rho_S(p))$ in Fig.3 (blue line). It should be noted
that the quantum state $\ket{\Phi_6}=U_{EF}\ket{\Psi_6}$ is also the
purification of state $\rho_S(p)$, in which the unitary operation
affects the concurrence $C_{EF}$ and the multipartite entanglement
$E_{ms}$. However, it does not change our conclusion.

\emph{Discussion and conclusion--} For the $N$-qubit case, there
also exists the entangled mixed state in which the $k$-tangles
entanglement are all vanished( here, $k=2,3,...,N$). Because
$\rho_{N}$ is a mixed state, the purified state
$\ket{\Psi}_{N\mathcal{E}}$ is entangled in the partition
$N|\mathcal{E}$. The entanglement $E_{N|\mathcal{E}}$ has a
hierarchy structure, in which, after a POVM on environment
$\mathcal{E}$, the genuine multiqubit QCR entanglements
$\{E_{A_1Ai\mathcal{E}},
E_{A_1A_iA_j\mathcal{E}},...,E_{A_1A_2,...,A_N\mathcal{E}}\}$ reduce
to the bipartite entanglement measured by mixed state one-tangle
$\tau_{A_1}$. As an example, we consider a quantum state given by
\begin{equation}\label{11}
    \rho_{A_1A_2...A_N}=\alpha \proj{1^{\otimes N}}+(1-\alpha)\proj{W_N},
\end{equation}
where $\ket{W_N}$ is an $N$-qubit $W$ state with the parameter
chosen as $\alpha=1/(N+1)$. The quantum state is invariant under
qubit permutation, and there is only one independent two-qubit
density matrix $\rho_{A_iA_j}$. After a simple derivation, one can
obtain $C_{ij}=0$. Furthermore, one can deduce the higher tangles
$\tau_k(\rho_{N})=0$ for $k=3,4,...,N$, which is because the quantum
state can be written as the mix of a product state and a $W$ state.
However, the mixed state one-tangle is
$\tau_{A_{1}}(\rho_N)=4(N-1)/(N^2+N)$, which means that the mixed
state $\rho_{A_1A_2...A_N}$ is entangled. According to the
purification theorem, its purified state has the form
$\ket{\Psi_{N+1}}=\sqrt{\alpha}\ket{1^{\otimes
N+1}}+\sqrt{1-\alpha}\ket{W_N}\otimes\ket{0}$. After a POVM on the
environment system $\rho_{A_{N+1}}$, the genuine multiqubit QCR
entanglements between system and environment reduces to the
one-tangle in the partition $A_1|A_2...A_N$.

We have chosen the measure $E_{ms}$ to quantify the total genuine
multi-qubit entanglement in a pure state. When the mixed state
higher tangles $\tau_k$ ($k>2$) vanish, they are compatible with the
$E_{ms}$. While we need to consider their compatibility whenever
these tangles are nonzero, the analytical solutions for higher
tangles of mixed state are still awaited, especially for the
nonzero-tangle case.

In conclusion, with the help of the QCRs and a purification
scenario, we have analyzed the one-tangle (or negativity) in a class
of multi-qubit mixed states without multipartite tangles. It has
been found that the entanglement in the purified state plays an
important role. Especially, whenever the mixed state has no
concurrence and higher tangles, its entanglement is just a kind of
the genuine multipartite QCR-entanglement between the mixed state
system and its environment.

\emph{Acknowledgments.--} The authors would like to thank G.-P. He,
M. Yang and Z.-Y. Xue for many helpful discussions. The work was
supported by the RGC of Hong Kong under  Grants No. HKU7051/06P,
HKU7044/08P, and HKU-3/05C, the URC fund of HKU, and NSF China Grant
No. 10429401. Y. K. B. was also supported by the fund of Hebei
Normal University. M.Y.Y. was also supported by the Foundation for
Universities in Fujian Province (Grant No. 2007F5041) and NSF-China
Grant No. 60878059.

\end{document}